\documentclass[12pt]{article}
\usepackage[dvips]{color}
\usepackage{amsmath}
\usepackage{graphicx}
\usepackage{subfigure}
\usepackage{epsfig}
\usepackage{amsmath}
\usepackage{cite}
\usepackage{color}

\begin{document}
\begin{center}
\Large{\bf Investigation of thermal properties of Hulth\'{e}n potential from statistical and superstatistical perspectives with various distribution}\\
 \small \vspace{1cm}
 {\bf Amir Hossein Khorram manesh $^{\star}$\footnote {Email:~~~a.khoram02@umail.umz.ac.ir}}, \quad
 \small \vspace{0.3cm} {\bf J. Sadeghi $^{\star}$\footnote {Email:~~~pouriya@ipm.ir,\quad	j.sadeghi@umz.ac.ir}}\quad
 {\bf Saeed Noori Gashti$^{\dagger}$\footnote {Email:~~~saeed.noorigashti@stu.umz.ac.ir}}\quad\\
\vspace{0.5cm}$^{\star}${Department of Physics, Faculty of Basic
Sciences,\\
University of Mazandaran
P. O. Box 47416-95447, Babolsar, Iran}\\
\vspace{0.5cm}$^{\dagger}${School of Physics, Damghan University, Damghan, 3671641167, Iran.}\\
\small \vspace{1cm}
\end{center}
\begin{abstract}
The Hulth\'{e}n potential is a short-range potential that has been widely used in various fields of physics. In this paper, we investigate the distribution functions for the Hulth\'{e}n potential by using statistical and superstatistical methods. We first review the ordinary statistics and superstatistics methods. We then consider some distribution functions, such as uniform, 2-level, gamma, and log-normal and F distributions. Finally, we investigate the behavior of the Hulth\'{e}n potential for statistical and superstatistical methods and compare the results with each other. We use the Tsallis statistics of the superstatistical system. We conclude that the Tsallis behavior of different distribution functions for the Hulth\'{e}n potential exhibits better results than the statistical method. We examined the thermal properties of the Hulth\'{e}n potential for five different distributions: Uniform, 2-level, Gamma, Log-normal, and F. We plotted the Helmholtz free energy and the entropy as functions of temperature for various values of q. It shows that the two uniform and 2-level distributions have the same results due to the universal relationship and that the F distribution does not become ordinary statistics at q=1. It also reveals that the curves of the Helmholtz free energy and the entropy change their order and behavior as q increases and that some distributions disappear or coincide at certain values of q. One can discuss the physical implications of our results and their applications in nuclear and atomic physics in the future.
\end{abstract}
\tableofcontents
\section{Introduction}
A specific instance of the Eckart potential is the Hulth\'{e}n potential, which is a notable short-range potential in physics\cite{1,2}. This potential has been applied across various fields, including nuclear and atomic physics. The Hulth\'{e}n potential demonstrates Coulombic behavior for small values of \(r\) and decays exponentially for large values of \(r\)\cite{3,4,5}. Generally, potentials can be solved either analytically and exactly or approximately and numerically. Several methods exist for solving potentials, such as the Asymptotic Iteration Method (AIM), the Nikiforov-Uvarov (NU) method, Ansatz, and Supersymmetric (SUSY) approaches. Supersymmetric Quantum Mechanics (SUSYQM) involves key concepts like factorization, Hamiltonian hierarchy, and shape invariance, which assist in solving potentials\cite{6}. For states with zero angular momentum, the Hulth\'{e}n potential in the radial Schr$\ddot{o}$dinger equation has an analytical solution. For non-zero angular momentum states, numerical methods are used to find the eigenstates. Researchers have applied the Hamiltonian hierarchy problem within the SUSYQM framework and found that the associated supersymmetric partner potentials simulate the effect of the centrifugal barrier. Using first-order perturbation theory, they obtained the analytical solution of the effective Hulth\'{e}n potential for states with non-zero angular momentum\cite{7}. Researchers have also obtained the eigenspectrum of energy and momentum for the time-independent and time-dependent Hulth\'{e}n-screened cosine Kratzer potential by solving the Schr$\ddot{o}$dinger equation in D dimensions using the Qiang-Dong proper quantization rule and the SUSYQM method\cite{8}. The modified radial Schr$\ddot{o}$dinger equation can be solved analytically by two methods: the NU method in ordinary quantum mechanics and the shape invariance principle in SUSYQM. Researchers obtained the bound state solution of the modified radial Schr$\ddot{o}$dinger equation for the Manning-Rosen plus Hulth\'{e}n potential using both the NU and SUSYQM methods, demonstrating equivalent expressions for the energy eigenvalues and transformations of the radial wave functions\cite{9}. Additionally, the analytical solution of the radial Schr$\ddot{o}$dinger equation for the Hulth\'{e}n potential was obtained using the NU method and SUSYQM approaches. The energy levels and corresponding normalized eigenfunctions were worked out in terms of orthogonal polynomials for arbitrary angular momentum states\cite{10}. Researchers using the conventional NU method obtained the solution of the Schr$\ddot{o}$dinger equation for the Hulth\'{e}n potential in D-dimensions with an exponential approximation of the centrifugal term\cite{11}. The exact solution of the radial Schr$\ddot{o}$dinger equation for the Eckart plus Hulth\'{e}n potential was obtained using the NU method, and the energy eigenvalue of the Hulth\'{e}n potential was computed\cite{12,13}. An approximate solution of the Schr$\ddot{o}$dinger equation for the generalized Hulth\'{e}n potential with non-zero angular momentum was presented via the NU method, with bound state energy eigenvalues and eigenfunctions obtained in terms of Jacobi polynomials\cite{14}. Recent works using the generalized pseudospectral method calculated the energy levels, transition oscillator strengths, and multipole polarizabilities for the hydrogen-like atom with the Hulth\'{e}n potential. They also studied various information-theoretic measures and the confined hydrogen atom in the modified Hulth\'{e}n potential, finding interesting entropy patterns in different combinations of principal quantum number and orbital quantum number\cite{a,b}.\\
Supersymmetry was originally proposed in relativistic quantum field theory as an extension of poincar\'{e} symmetry. In 1976, Herman Nikolai applied a similar extension to non-relativistic quantum mechanics. Witten also presented a one-dimensional model in 1981, and supersymmetry emerged as a key concept in many fields of Physics such as quantum mechanics, mathematical physics, statistical mechanics, atomic physics, nuclear physics, and condensed matter physics\cite{15,16}.\\\\
SUSYQM helped to solve problems of non-relativistic quantum mechanics in such a way that we can now solve, exactly by analytical approaches, solvable potentials, and by various numerical methods, potentials that are not solvable. We use the concept of shape invariance, partner potentials, partner Hamiltonian, superpotential, and the relation between eigenfunctions and eigenvalues to solve the Schr\"{o}dinger equation for a potential\cite{16}.
The factorization method is fascinating in SUSY QM for solving differential equations. The Hamiltonian formalism in supersymmetric quantum mechanics reveals that a Hamiltonian can be split into two operators, and with this method, we can solve first-order equations instead of a quadratic equation.
The Factorization method was first proposed by Darboux and then used in quantum mechanics by Schr\"{o}dinger\cite{17,18}. Infeld and Hull studied a wide range of second-order equations in \cite{19}.
One of the main objectives in quantum mechanics since its inception is to find the solution for the Schr\"{o}dinger equation to obtain eigenfunctions and eigenvalues of a potential.\\\\
The Hulth\'{e}n potential that simulates the centrifugal barrier for any angular momentum state was used to analyze the time-independent Schr\"{o}dinger equation and they showed that the eigenvalues and eigenfunctions of the generalized Hulth\'{e}n potential with non-zero angular momentum can be obtained approximately in terms of Jacobi polynomials\cite{4}. In this paper, we adopt the generalized Hulth\'{e}n potential from \cite{20},
\begin{equation}\label{1}
  V(r)=\frac{V_{0}}{1-\exp(Ar)}+\frac{2A^{2}\exp(Ar)}{(1-\exp(Ar))^{2}}.
\end{equation}
By the factorization method, the energy eigenvalues are obtained from the potential, where R is the potential radius, $V_{0}$ is a positive constant, $A$ is the screening parameter, and $V_{0}>A^{2}$. The Schr\"{o}dinger equation was factorized by Raising and Lowering Operators and the analytical solution of the generalized Hulth\'{e}n potential in zero angular momentum state was derived \cite{4}. This energy eigenvalue will be used to compare the thermal properties in ordinary statistical mechanics and superstatistics,
\begin{equation}\label{2}
  E_{m}(\alpha)=-\frac{\hbar^{2}}{8Ma^{2}}(\alpha+m)^{2}.
\end{equation}
The potential depends on the parameters $ M $, $ a $, $ m $, and $\alpha$, where $ M $ is mass, $ a $ is Bohr radius, m is a non-negative integer, which is an index related to several Jacobi terms and $\alpha$ is a real parameter. In various fields of physics, the Hulth\'{e}n potential is a short-range potential that is widely used. In this paper, we apply statistical and superstatistical methods to study the distribution functions for the Hulth\'{e}n potential. We then investigate some distribution functions, such as uniform, 2-level, gamma, log-normal, and F distributions. We also analyze the behavior of the Hulth\'{e}n potential for statistical and superstatistical methods and compare the outcomes with each other. We consider the Tsallis statistics of the superstatistical system. Tsallis statistics is a generalization of the usual Boltzmann–Gibbs statistics, which measures the disorder or uncertainty of a system. Tsallis statistics can describe systems that are out of equilibrium or have long-range interactions or memory effects. Tsallis statistics has been used to study dark energy, which is the mysterious force that drives the accelerated expansion of the universe. One way to study dark energy is to use the holographic principle, which relates the entropy of a system to its boundary area. By using Tsallis entropy instead of Boltzmann–Gibbs entropy, one can obtain different models of holographic dark energy, which have different properties and behaviors. Some of these models can explain the current observations of the universe better than others. Tsallis statistics and holographic dark energy are active areas of research in physics and cosmology\cite{100,101,102,103,104}\\\\ Based on the above concepts, this article is organized as follows: Section 2 introduces an overview of ordinary statistics and superstatistics methods. In section 3, we use the thermal properties with respect to the mentioned methods to study the distribution functions for the Hulth\'{e}n potential. We also analyze the behavior of the Hulth\'{e}n potential for statistical and superstatistical methods and compare the outcomes with each other. We discuss the results that we derived from the connection between the concepts that we introduced in detail in section 4. We end our paper by giving a summary of our findings in section 5.
\section{Overview of ordinary statistics and superstatistics}
To obtain the statistical properties of the Hulth\'{e}n potential, we need the partition function. In ordinary statistical mechanics, we have the following relation,
\begin{equation}
  Z=\sum_{n=0}^{\infty}\exp(-\beta E_{n})=\int_0^\infty \exp(-\beta E)dE,
\end{equation}
where $E$ is the energy of a microstate associated with each cell of our system and $\beta \equiv \frac{1}{K_{b} T}$ is the thermodynamic beta, with $K_{b}$ being the Boltzmann constant and $T$ the temperature. The partition function in superstatistics is different. We first need to define the generalized Boltzmann factor as follows,
\begin{equation}
  B(E)=\int_0^\infty f(\beta) \exp(-\beta E) d\beta,
\end{equation}
where $f(\beta)$ is the probability density function of $\beta$. We can now express the partition function in superstatistics,
\begin{equation}
  Z=\int_0^\infty B(E)dE.
\end{equation}
These relations are derived from \cite{21,22}. The partition function was first introduced by Boltzmann. This function can be expressed in terms of $\beta$ or $T$. Other statistical representations were proposed by Gibbs, Einstein, Boltzmann-Gibbs, and Tsallis and finally, we have the super statistics representation, which was originally conceived by Wilk and E.g. Woldarczyk and then reformulated by Beck and Cohen\cite{23,24}.\\
Superstatistics deals with non-equilibrium systems that have complex dynamics with large fluctuations of intensive quantities, such as inverse temperature, chemical potential, or energy dissipation, on long-time scales. The name of superstatistics comes from the superposition of two statistics, namely the superposition of $\beta$ and the ordinary Boltzmann factor $\exp(-\beta E)$. One statistic is the ordinary statistics related to the $\beta$ that is approximately constant in spatial regions, i.e. the ordinary Boltzmann factor $\exp(-\beta E)$, where $E$ is the energy of each spatial region. The other statistic is described by the long time averaging over the fluctuating $\beta$. In the superstatistics formulation, we define a Tsallis parameter q. For q = 1, the generalized Boltzmann factor reduces to the ordinary Boltzmann factor\cite{21,22}.\\\\
We can distinguish two types of superstatistics: Tsallis and Kaniadakis. In this paper, we will study the thermal properties of the Hulth\'{e}n potential for Tsallis superstatistics and compare it with ordinary statistical mechanics. For the mean value (expected value) of $\beta$ and $\beta^{2}$, one can write,
\begin{equation}
  <\beta>= \int_0^\infty \beta f(\beta) d\beta,
\end{equation}
\begin{equation}
  <\beta^{2}>= \int_0^\infty \beta^{2} f(\beta) d\beta,
\end{equation}
where $<\beta^{2}>=\beta_{0}^{2}$ and for variance we have,
\begin{equation}
  \sigma^2=<\beta^{2}>+<\beta>^{2}.
\end{equation}
In this section, we will present some distributions and then demonstrate a universal relation for all of them\cite{21,22}. By using this universal relation, we will derive the thermal properties of these distributions and plot them for comparison.

In statistics, the uniform distribution is a member of a family of symmetric statistical distributions that have a constant value for the distribution function in a certain range. The uniform distribution is often used to model situations where all outcomes are equally likely, such as rolling a fair die or picking a random card from a deck. A 2-level distribution can be related to systems that have variable intensity parameter values discretely and with the same probability. For example, a Brownian particle moves between two different states with two different friction constants or two different values of chemical potential. A 2-level distribution is a special case of a discrete uniform distribution, where the number of possible values is two. The gamma distribution in statistics is a distribution with two positive parameters, one parameter is related to the shape and the other to the scale. The gamma distribution has many applications throughout science. The gamma distribution can describe the waiting time between Poisson events, among other examples. A log-normal distribution is a continuous probability distribution of a random variable where the logarithm is normally distributed. For example, if the random variable $x$ has a log-normal distribution, then $y=\ln(x)$ has a normal distribution. Also, if $y$ has a normal distribution, then $x=\exp(y)$ has a log-normal distribution. Log-normally distributed random variables take only positive real values. The log-normal distribution can model phenomena that are multiplicative. The F distribution is a continuous probability distribution that has a range of non-negative values. This distribution is the ratio of two independent chi-square distributions, each of which was divided by the degrees of freedom. This distribution has two parameters, one parameter is the degrees of freedom of the numerator of the fraction, and the other is related to the degrees of freedom of the denominator of the fraction. The F distribution has many uses, including variance analysis, and regression analysis. The F distribution can compare the variability of two sets of data, test the significance of regression coefficients, and assess the goodness of fit of a model.
\subsection{Uniform distribution}
We begin with the Uniform distribution of $\beta$, which is a simple model. Its distribution function for $0 \leq a \leq \beta \leq a+b$ is,
\begin{equation}
  f(\beta) = \frac{1}{b}.
\end{equation}
For the mean value and the variance of $\beta$, we will have,
\begin{eqnarray} \nonumber
<\beta> &=& a+\frac{b}{2} \\
\sigma^{2} &=& \frac{b^{2}}{12}.
\end{eqnarray}
\subsection{2-Level distribution}
This distribution models the subsystems that can alternate between two distinct discrete values of the fluctuating intensive parameter with equal likelihood. The 2-level distribution function is as follows,
\begin{equation}
  f(\beta) = \frac{\delta(a)}{2}+\frac{\delta(a+b)}{2}.
\end{equation}
Also, the mean value and variance is given by,
\begin{eqnarray} \nonumber
<\beta> &=& a+\frac{b}{2} \\
\sigma^{2} &=& \frac{b^{2}}{4}.
\end{eqnarray}
\subsection{Gamma distribution}
The Gamma distribution of the inverse temperature $\beta$ results in Tsallis statistics, which is the most prominent example of superstatistics so far. The distribution function is given by,
 \begin{equation}
   f(\beta) = \frac{1}{b\Gamma(c)}(\frac{\beta}{b})^{c-1} e^{(\frac{-\beta}{b})},
 \end{equation}
where $c > 0$ and $b > 0$. The mean value and the variance are given by,
 \begin{eqnarray} \nonumber
 <\beta> &=& bc \\
\sigma^{2} &=& b^{2} c.
 \end{eqnarray}
\subsection{Log-normal distribution}
The corresponding distribution function is as follows,
 \begin{equation}
   f(\beta)= \frac{1}{\beta s \sqrt{2 \pi}} \exp\frac{-(\log(\frac{\beta}{m}))^{2}}{2 s^{2}},
 \end{equation}
where $ m $ and $ s $ are free parameters Corresponding to the Log-normal distribution. The mean and variance of $ \beta $ defined as follows, respectively,
 \begin{eqnarray} \nonumber
<\beta> &=& m\sqrt{\omega} \\
\sigma^{2} &=& m^{2} \omega (\omega-1).
\end{eqnarray}
The quantity $ \omega $ is equal to the exponential of the square of $ s $, i.e., $ \omega = e^{s^{2}} $.
\subsection{F-distribution}
The distribution function for $ \beta \in [0,\infty] $ is given by the following expression
\begin{equation}
   f(\beta)= \frac{\Gamma(\frac{\nu + \omega}{2})}{\Gamma(\frac{\nu}{2})\Gamma(\frac{\omega}{2})} (\frac{b\nu}{\omega})^{\frac{\nu}{2}} \frac{\beta^{\frac{\nu}{2}-1}}{(1+\frac{\nu \beta b}{\omega})^{\frac{\nu+\omega}{2}}},\\
\end{equation}
where $ \nu $, $ \omega $ and $b$ are positive integers parameters. The mean and variance of $ \beta $ are calculated as follows
\begin{eqnarray}  \nonumber
 <\beta> &=& \frac{\omega}{b(\omega-2)} \\
\sigma^{2} &=& \frac{2 \omega^{2} (\nu+\omega-2)}{{b^{2} \nu (\omega-2)^{2} (\omega-4)}}.
\end{eqnarray}
In this paper, we set $ \nu = 4 $
\subsection{Universal relation for distributions}
The following relations can be applied to any distribution. So the generalized Boltzmann factor is as follows,
\begin{equation}
  B(E)=\exp(-\beta_{0}E)(1 + \frac{1}{2} \sigma^{2} E^{2} + \sum_{r=3}^{\infty} \frac{(-1)^{r}}{r!} <(\beta - \beta_{0} )^{r} E^{r}>),
\end{equation}
where
\begin{eqnarray}
 \nonumber
  <(\beta - \beta_{0})^{r}> &=& \sum_{0}^{r} \binom{r}{j} <\beta^{j}> (-\beta_{0})^{r-j}  \\
   \sigma^{2} &=& (q-1) (\beta_{0})^{2}  \\  \nonumber
  q &=& \frac{<\beta^{2}>}{<\beta>^{2}}.
\end{eqnarray}
If the generalized Boltzmann factor is expressed in terms of $q$ and $\beta_{0}$, then we obtain
\begin{equation}
  B(E)= \exp(-\beta_{0} E) (1+ (q-1) \beta_{0}^{2} E^{2} + g(q) \beta_{0}^{3} E^{3} +...),
\end{equation}
where $ g(q) $ is a function that varies depending on the distribution. For the Uniform and 2-level distributions, $ g(q) $ is considered as follows,
\begin{equation}
  g(q)=0.
\end{equation}
For the Gamma distribution, $ g(q) $ is as,
\begin{equation}
  g(q)=- \frac{1}{3} (q-1)^{2}.
\end{equation}
So for the Log-normal distribution, we have,
\begin{equation}
  g(q)= - \frac{1}{6} (q^{3}- 3q +2).
\end{equation}
Also, for the F-distribution with $ \nu=4 $, $g(q)$ is,
\begin{equation}
  g(q)= - \frac{1}{3} \frac{(q-1) (5q-6)}{3-q}
\end{equation}
When $q=1$, the generalized Boltzmann factors for all distributions are reduced to the ordinary Boltzmann factor of statistical mechanics. For small fluctuations of $ \beta $, all distributions behave similarly and their Boltzmann factors are comparable. However, for large fluctuations, we can differentiate between them and observe that the Boltzmann factors of each distribution have distinct terms of third order (and higher).
In the next section, we will compute the partition functions and compare the thermal properties of the Hulth\'{e}n potential in ordinary statistics and superstatistics.
\section{Thermal properties}
Our goal is to derive the thermal properties of the Hulth\'{e}n potential in both ordinary statistics and superstatistics. To do this, we first need to compute the partition function and then apply the following relations to obtain the thermal properties.
\subsection{Thermal Relations}
According to the points mentioned above, we introduce the following functions to check the thermal properties. Hence we will have,
\begin{equation}
Z=\int_0^\infty B(E)dE.
\end{equation}
The above equation is the partition function. Now we will have some relations for the internal energy, Helmholtz free energy, and entropy, respectively,
\begin{equation}
U=-\frac{\partial}{\partial \beta} \ln(Z)= K_{b} T^{2} \frac{\partial \ln(Z)}{\partial T},
\end{equation}
\begin{equation}
A=-\frac{1}{\beta} \ln(Z) = - K_{b} T \ln(Z),
\end{equation}
and
\begin{equation}
S=K_{b} \beta U + K_{b} \ln(Z)=K_{b} T \frac{\partial \ln(Z)}{\partial T} + K_{b} \ln(Z).
\end{equation}
\subsection{Ordinary Statistical Mechanics}
By substituting the energy eigenvalue of the generalized Hulth\'{e}n potential into the partition function, we can derive the thermal properties as follows,
\begin{eqnarray}
\nonumber
Z &=& \frac{1}{2} \sqrt{\frac{8 \pi M a^{2} K_{b} T}{\hbar^{2}}} \\ \nonumber
U &=& \frac{1}{2}  K_{b} T  \\
A &=& - K_{b} T \ln(\frac{1}{2} \sqrt{\frac{8 \pi M a^{2} K_{b} T}{\hbar^{2}}}) \\ \nonumber
S &=& K_{b} \ln(\frac{1}{2} \sqrt{\frac{8 \pi M a^{2} K_{b} T}{\hbar^{2}}}) + \frac{K_{b}}{2}. \\  \nonumber
\end{eqnarray}

\subsection{Uniform and 2-Level distributions}
The Uniform and 2-Level distributions share the same generalized Boltzmann factor, which implies that the first two terms of their universal relation of the generalized Boltzmann factor are identical. Therefore, we obtain,
 \begin{eqnarray}
\nonumber
Z &=& \frac{5+3q}{16} \sqrt{\frac{8 \pi M a^{2} K_{b} T}{\hbar^{2}}} \\ \nonumber
U &=& \frac{1}{2}  K_{b} T  \\
A &=& -K_{b} T \ln(\frac{5+3q}{16} \sqrt{\frac{8 \pi M a^{2} K_{b} T}{\hbar^{2}}}) \\ \nonumber
S &=& K_{b} \ln(\frac{5+3q}{16} \sqrt{\frac{8 \pi M a^{2} K_{b} T}{\hbar^{2}}}) + \frac{K_{b}}{2}. \\ \nonumber
\end{eqnarray}
It is evident that all relations reduce to ordinary statistics when $q=1$
\subsection{Gamma distribution}
For the Gamma distribution, we obtain,
  \begin{eqnarray}
\nonumber
Z &=& \frac{-15 q^{2} + 39q}{48} \sqrt{\frac{8 \pi M a^{2} K_{b} T}{\hbar^{2}}} \\ \nonumber
U &=& \frac{1}{2}  K_{b} T  \\
A &=& -K_{b} T \ln(\frac{-15 q^{2} + 39q}{48} \sqrt{\frac{8 \pi M a^{2} K_{b} T}{\hbar^{2}}}) \\ \nonumber
S &=& K_{b} \ln(\frac{-15 q^{2} + 39q}{48} \sqrt{\frac{8 \pi M a^{2} K_{b} T}{\hbar^{2}}}) + \frac{K_{b}}{2} \\ \nonumber
\end{eqnarray}
In this case, the relations also reduce to ordinary statistics when $q=1$.
\subsection{Log-normal distribution}
The thermal properties of the Log-normal distribution are as follows,
  \begin{eqnarray}
\nonumber
Z &=& \frac{-15 q^{2} + 63q}{96} \sqrt{\frac{8 \pi M a^{2} K_{b} T}{\hbar^{2}}} \\ \nonumber
U &=& \frac{1}{2}  K_{b} T  \\
A &=& -K_{b} T \ln(\frac{-15 q^{2} + 63q}{96} \sqrt{\frac{8 \pi M a^{2} K_{b} T}{\hbar^{2}}}) \\ \nonumber
S &=& K_{b} \ln(\frac{-15 q^{2} + 63q}{96} \sqrt{\frac{8 \pi M a^{2} K_{b} T}{\hbar^{2}}}) + \frac{K_{b}}{2} \\  \nonumber
\end{eqnarray}
Similarly, the relations reduce to ordinary statistics when $q=1$ for this case.

\subsection{F-distribution}
As we stated in this paper, we assumed that $\nu=4$ for the F-distribution, which leads to a special case of it. Hence, we have,
 \begin{eqnarray}
\nonumber
Z &=& \frac{\frac{7}{4} q^{2} - 4q + \frac{15}{8}}{q-3} \sqrt{\frac{8 \pi M a^{2} K_{b} T}{\hbar^{2}}} \\ \nonumber
U &=& \frac{1}{2}  K_{b} T  \\
A &=& -K_{b} T \ln(\frac{\frac{7}{4} q^{2} - 4q + \frac{15}{8}}{q-3} \sqrt{\frac{8 \pi M a^{2} K_{b} T}{\hbar^{2}}}) \\ \nonumber
S &=& K_{b} \ln(\frac{\frac{7}{4} q^{2} - 4q + \frac{15}{8}}{q-3} \sqrt{\frac{8 \pi M a^{2} K_{b} T}{\hbar^{2}}}) + \frac{K_{b}}{2}. \\  \nonumber
\end{eqnarray}
In this special case of the F-distribution, the relations do not reduce to ordinary statistics when $q=1$.

\section{Discussion $\&$ result}
In all Figs related to the thermal properties of the Hulth\'{e}n potential, i.e. Helmholtz free energy and entropy diagrams, we examined the following distributions:
Uniform, 2-level, Gamma, lognormal and F.
Note: the F distribution in this article is in a special case and all distributions are taken from \cite{21}
In all the Figs, due to the universal relationship that we mentioned, the two uniform and 2-level distributions have the same results and are consistent with each other. They have the same Boltzmann factor.
Another point is that, unlike all the above distributions, the F distribution does not become ordinary statistics at q=1, while all other distributions exactly become ordinary statistics at q=1, that is, the distribution function, the Helmholtz free energy, and the entropy become the same except for F.
In q=1, the diagram for both the Helmholtz free energy and entropy, all mentioned distributions coincide except for the F distribution.
At q=1.1, the difference between the distributions becomes apparent, that is, they are slightly separated and distinguishable from each other, but they are still close to each other.
Also, as q increases, these curves are further apart.
For both the Helmholtz free energy and the entropy at q=2, uniform, and log-normal distributions are found.
Another point that happened is that both in Helmholtz free energy and entropy before q=2, the curves of two uniform and log-normal distributions change.
That is, for example, in the diagrams of Helmholtz free energy and entropy from q=1 to q=2, the order of the curves is preserved, i.e. they are one after the other, but after q=2, the order is still preserved, but the two distributions, uniform and log-normal, are switched. For example, before q=2, it was first uniform, then log-normal, after that, first log-normal, then uniform. The arrangements for entropy and Helmholtz free energy are analogous.\\\\
Helmholtz free energy diagram in terms of temperature in Fig(1a): For the Helmholtz free energy at q=1, it can be seen that all distributions are equal to the ordinary statistics and their curves match except for the F distribution.
For q=1.1 in Fig (1b): the difference between the distributions begins. It can be said that gamma and uniform are close to each other. F distribution has a different behavior compared to other curves of other distribution functions.
For q=1.2 in Fig (2a):
It is in the same order as the previous diagram, but the curves become more open and F has a different behavior compared to other curves.
For q=1.5 in Fig (2b):
The gamma curve approaches the ordinary statistics, and the uniform and log-normal approach each other.
For q=1.7 in Fig (3a):
The F distribution disappears, and on the other hand, the curves become more open, gamma and normal statistics are closer to each other, and uniform and log-normal are closer to each other.
For q=1.8 in Fig (3b):
It is like the previous diagram.
For q=2 in Fig (4a):
Uniform and log-normal curves coincide and we do not have an F curve.
For q=2.1 in Fig (4b):
The uniform curve and the log-normal curve are separated from each other, with the difference that the place of their curves is shifted in the graph and we do not have the F curve.
For q=2.5 in Fig (5a):
We don't have an F curve, and the behavior of gamma is not similar to other curves, and the behavior of ordinary statistics is closer to log-normal and uniform.
For q=2.8 in Fig (5b):
Gamma and F distributions are determined.\\\\
Entropy diagram in terms of temperature for q=1 in Fig (6a):
Similar to the Helmholtz free energy, all the curves of the distributions coincide except for the F distribution.
For q=1.1 in Fig (6b):
The differences between the curves begin. It can be seen that the order of these curves is proportional to the order of the curves in Helmholtz's free energy. Uniform and gamma are very close to each other.
For q=1.2 in Fig (7a):
The curves are further apart and the F distribution behaves differently.
For q=1.5 in Fig (7b):
Gamma and normal statistics are close to each other and uniform, and on the other hand, log normal are also close to each other.
 For q=1.7 in Fig (8a):
  Gamma and normal are close together and above uniform and log-normal, which are close together, although the curves are opening.
  For q=1.8 in the Fig (8b):
It is similar to the previous diagram.
  For q=2 in Fig (9a):
Uniform and log-normal coincide and we don't have an F curve.
For q=2.1 in Fig (9b):
The uniform curve and the lognormal curve are separated, with the difference that their curves are switched in the graph and we do not have the F curve.
For q=2.5 in Fig (10a):
We don't have an F curve, and the behavior of gamma is not similar to other curves, and the behavior of normal statistics is closer to log-normal and uniform.
  For q=2.8 in Fig (10b):
F and Gamma do not have curves.
In conclusion, the analysis of different statistical distributions in connection with the Hulth\'{e}n potential, superstatistics, and normal statistics reveal intricate and varying behaviors under changing conditions. At \( q = 1 \), all distributions except the F distribution align with ordinary statistics, indicating classical statistical mechanics behavior. As \( q \) increases from 1.1 to 2.8, the differences between distributions become more pronounced.
The F distribution consistently exhibits distinct behavior compared to other distributions and even disappears entirely at certain values of \( q \) (e.g., \( q = 1.7 \), \( q = 2 \), \( q = 2.1 \), \( q = 2.5 \)). Gamma and Uniform distributions often behave similarly and are close to each other in many cases, while the Log-Normal distribution tends to coincide with the Uniform distribution at higher values of \( q \) but separates at \( q = 2.1 \).
Ordinary statistics often align more closely with the Log-Normal and Uniform distributions at higher values of \( q \). As \( q \) increases, the curves for both Helmholtz free energy and entropy become more open, indicating greater divergence in the behavior of different distributions. The disappearance of the F distribution at certain values of \( q \) suggests its behavior becomes significantly different or less relevant in these regimes.
The correlation between Helmholtz free energy and entropy is evident, as the order and behavior of the curves in the entropy diagrams are proportional to those in the Helmholtz free energy diagrams. This consistent relationship across different values of \( q \) highlights the complex and varying behavior of different statistical distributions, providing valuable insights into the thermodynamic properties of systems described by these distributions.
\section{Concluding remarks}
The Hulth\'{e}n potential is a short-range potential that has been widely used in various fields of physics. In this paper, we investigated some distribution functions for the Hulth\'{e}n potential by using statistical and superstatistical methods. We first reviewed the ordinary statistics and superstatistics methods, which were based on different assumptions about the fluctuations of the system parameters. We then considered some distribution functions, such as uniform, 2-level, gamma, log-normal, and F distributions, which were commonly used to model the superstatistical behavior of complex systems. Finally, we investigated the behavior of the Hulth\'{e}n potential for statistical and superstatistical methods and compared the results with each other. We used the Tsallis entropy of the superstatistical system, which was a generalization of the Boltzmann-Gibbs entropy and could capture the non-extensive and non-linear properties of the system. We concluded that the Tsallis behavior of different distribution functions for the Hulth\'{e}n potential exhibited better results than the statistical method, as it could account for the variations of the potential strength and the dimensionality of the system. We examined the thermal properties of the Hulth\'{e}n potential for five different distributions: Uniform, 2-level, Gamma, lognormal, and F. We plotted the Helmholtz free energy and the entropy as functions of temperature for various values of q. We found that the Uniform and 2-level distributions had the same results due to the universal relationship and that the F distribution did not become ordinary statistics at q=1. We also observed that the order and behavior of the curves changed as q increased and that some distributions disappeared or coincided at certain values of q. We used statistical tests to compare the distributions and to test our hypotheses. One could discuss the physical implications of our results and their applications in nuclear and atomic physics in the future, such as the stability and binding energy of the nuclei and the scattering and decay processes of the atoms.

\section{Appendix A: Figures of different distributions}
In this section, we have plotted the diagrams of Helmholtz free energy and entropy as functions of temperature for different values of the Tsallis parameter and for different distributions. The Helmholtz free energy is a thermodynamic potential that measures the useful work obtainable from a closed system at a constant temperature. The entropy is a measure of the disorder or randomness of a system.\\
In this article, we used a series of parameters that are applied equally in all graphs. The graphs include Helmholtz free energy and entropy as functions of temperature. In each graph, the curves related to normal statistics and superstatistical distributions are plotted side by side for different values of the parameter \( q \). The variable in these graphs is the temperature in Kelvin.
Now, we list the relevant parameters and units:\\
$\cdot$ Bohr radius \( a = 5.29 \times 10^{-11} \) m\\
$\cdot$ Boltzmann's constant \( K_b = 1.3806488 \times 10^{-23} \) J/K\\
$\cdot$ Planck's constant \( \hbar = 1.053472010 \times 10^{-34} \) J·s\\
$\cdot$ Atomic mass \( M = 1.66053906606 \times 10^{-27} \) kg\\
$\cdot$ \( 0 < n = m < \infty \)\\
$\cdot$ Parameter limits: \( 1 \rightarrow \beta \), \( 1 \rightarrow \alpha \)
\begin{figure}[h!]
 \begin{center}
 \subfigure[]{
 \includegraphics[height=15.5cm,width=8.5cm]{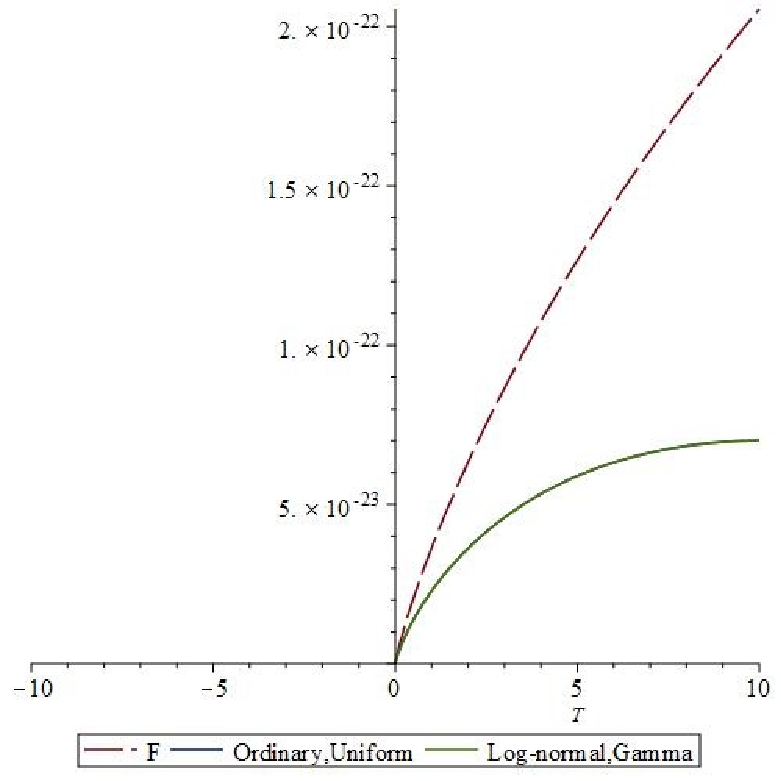}
 \label{1a}}
 \subfigure[]{
 \includegraphics[height=15.5cm,width=8.5cm]{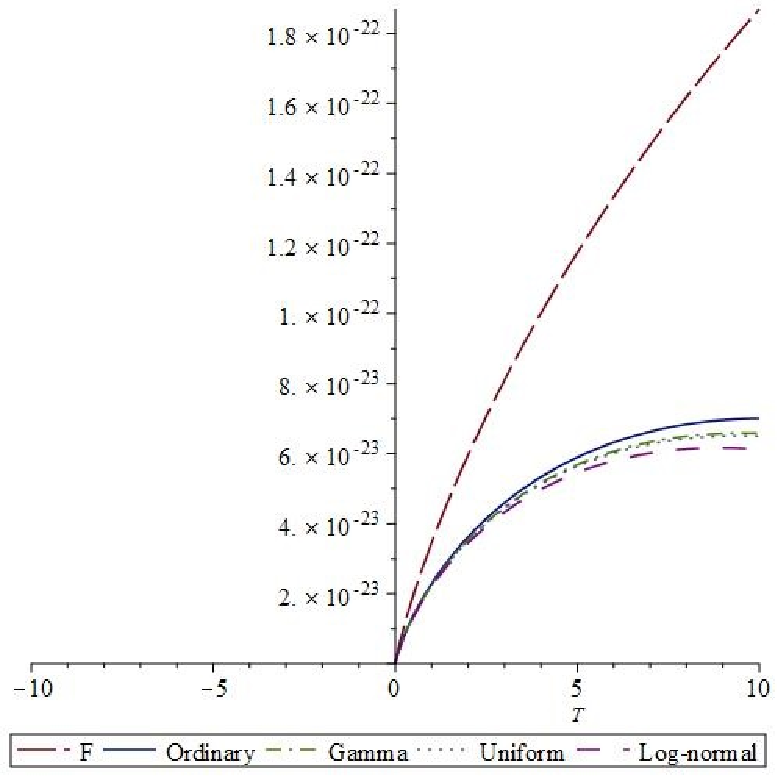}
 \label{1b}}
  \caption{\small{Helmholtz free energy in terms of temperature for different distributions with respect to $q=1, 1.1$ for figs 1a and 1b}}
 \label{1}
 \end{center}
 \end{figure}
\newpage
 \begin{figure}[h!]
 \begin{center}
 \subfigure[]{
 \includegraphics[height=15.5cm,width=8.5cm]{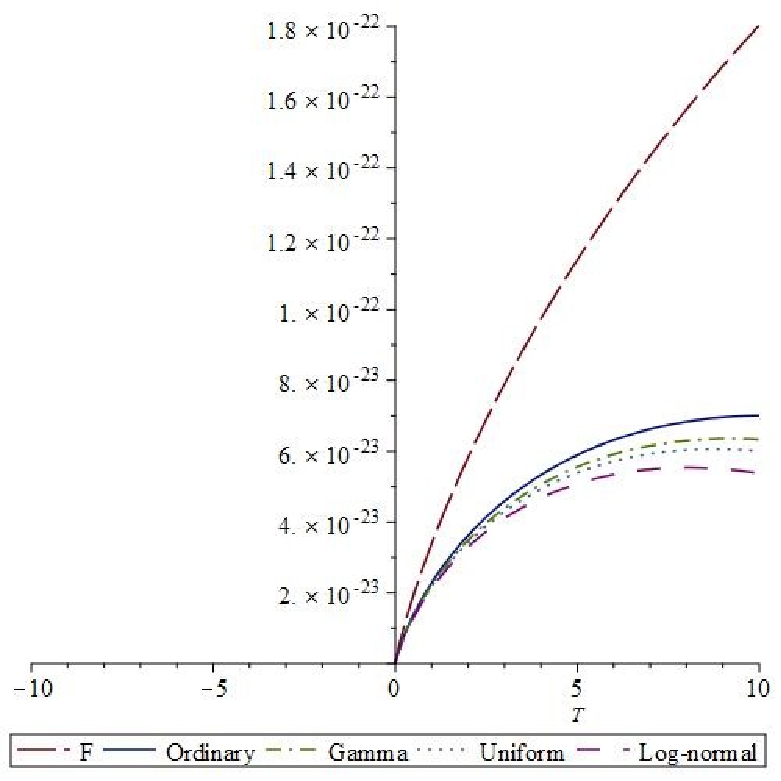}
 \label{2a}}
 \subfigure[]{
 \includegraphics[height=15.5cm,width=8.5cm]{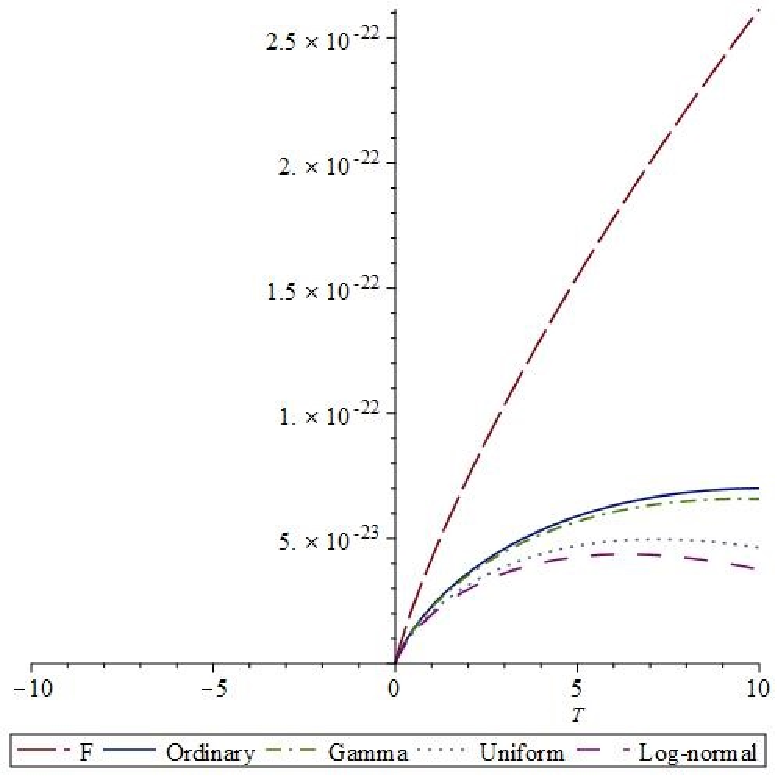}
 \label{2b}}
  \caption{\small{Helmholtz free energy in terms of temperature for different distributions with respect to $q=1.2, 1.5$ for figs 2a and 2b}}
 \label{2}
 \end{center}
 \end{figure}

  \begin{figure}[h!]
 \begin{center}
 \subfigure[]{
 \includegraphics[height=15.5cm,width=8.5cm]{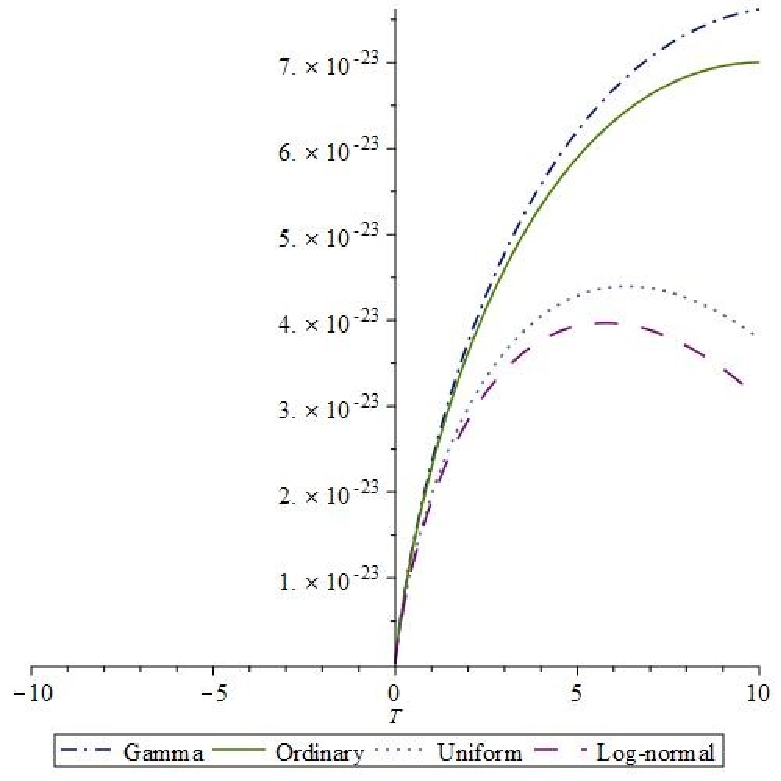}
 \label{3a}}
 \subfigure[]{
 \includegraphics[height=15.5cm,width=8.5cm]{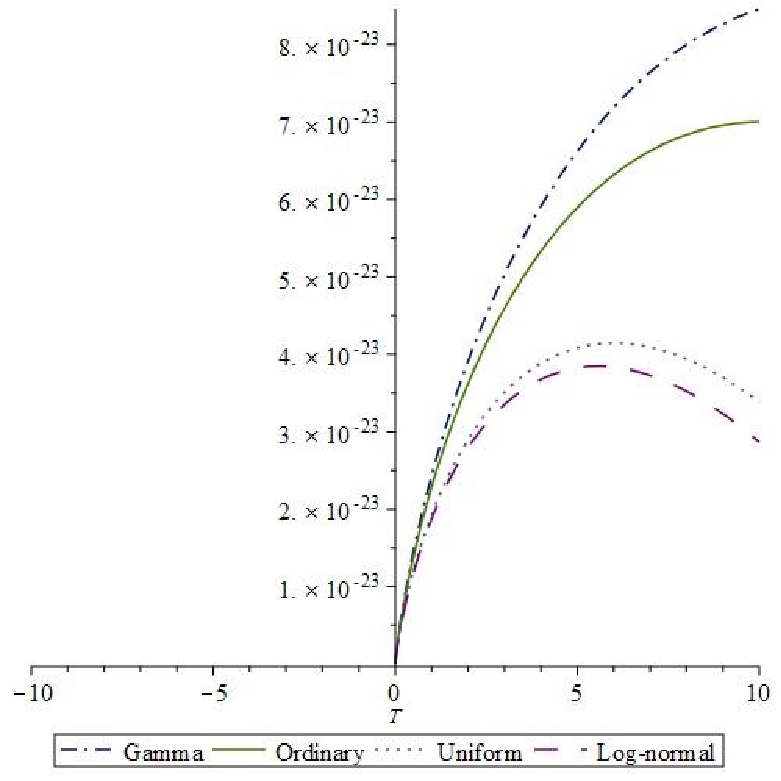}
 \label{3b}}
  \caption{\small{Helmholtz free energy in terms of temperature for different distributions with respect to $q=1.7, 1.8$ for figs 3a and 3b}}
 \label{3}
 \end{center}
 \end{figure}

   \begin{figure}[h!]
 \begin{center}
 \subfigure[]{
 \includegraphics[height=15.5cm,width=8.5cm]{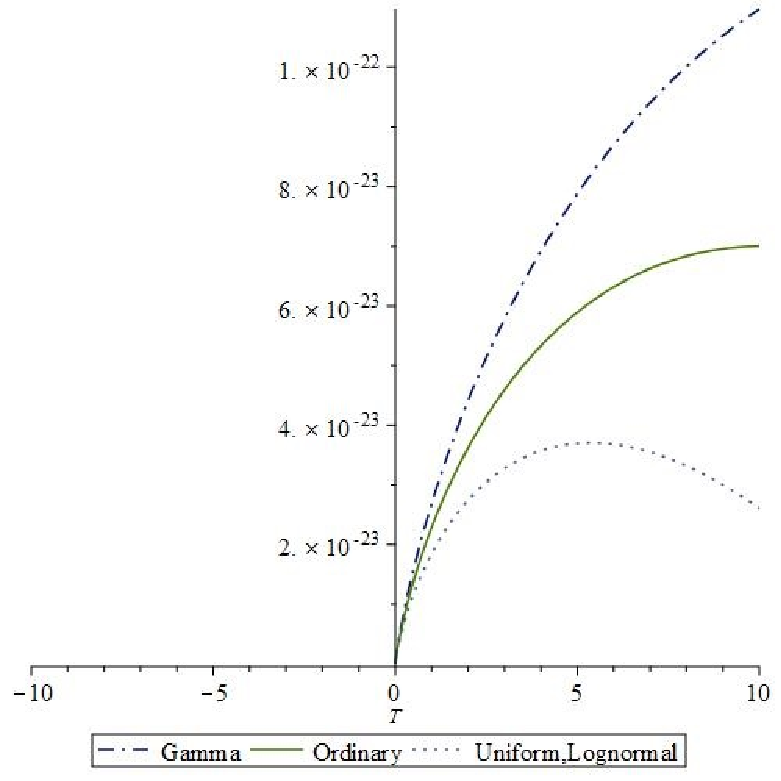}
 \label{4a}}
 \subfigure[]{
 \includegraphics[height=15.5cm,width=8.5cm]{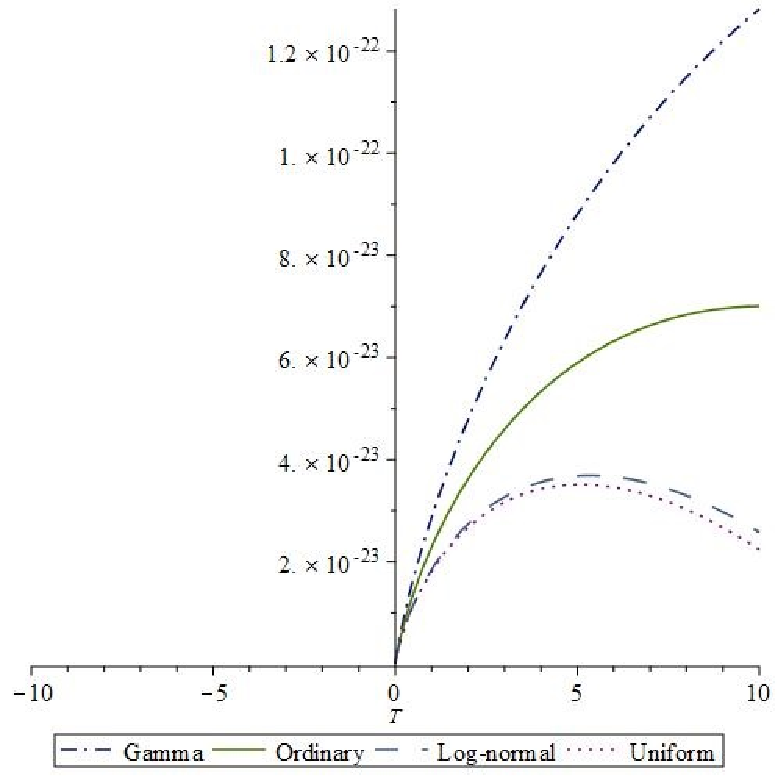}
 \label{4b}}
  \caption{\small{Helmholtz free energy in terms of temperature for different distributions with respect to $q=2, 2.1$ for figs 4a and 4b}}
 \label{4}
 \end{center}
 \end{figure}

   \begin{figure}[h!]
 \begin{center}
 \subfigure[]{
 \includegraphics[height=15.5cm,width=8.5cm]{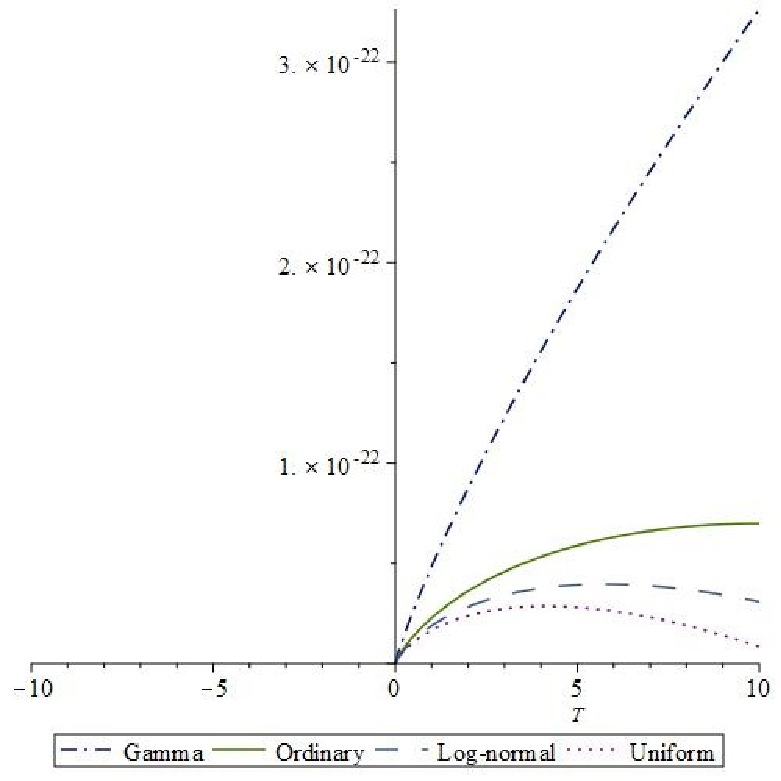}
 \label{5a}}
 \subfigure[]{
 \includegraphics[height=15.5cm,width=8.5cm]{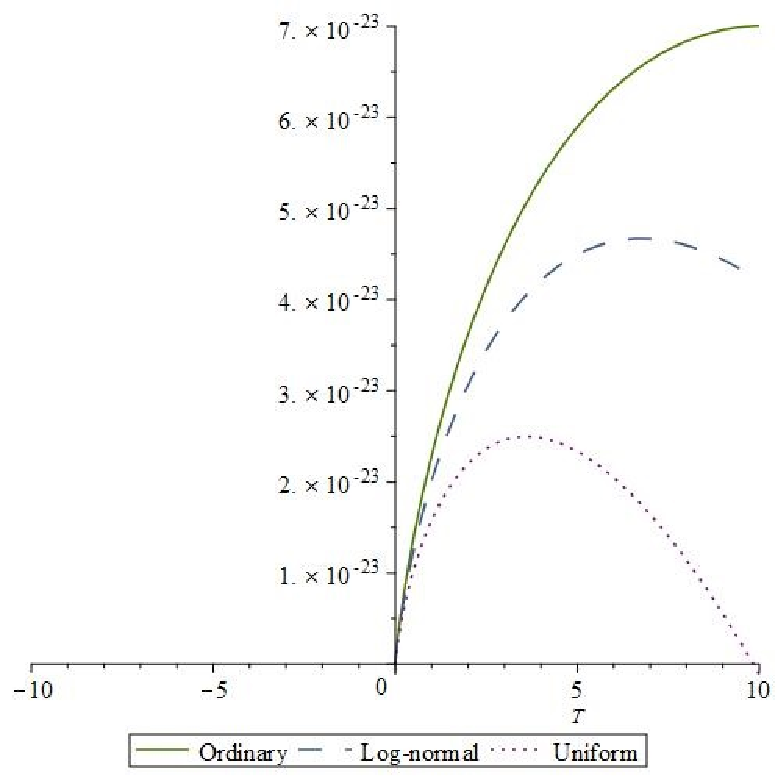}
 \label{5b}}
  \caption{\small{Helmholtz free energy in terms of temperature for different distributions with respect to $q=2.5, 2.8$ for figs 5a and 5b}}
 \label{5}
 \end{center}
 \end{figure}

\begin{figure}[h!]
 \begin{center}
 \subfigure[]{
 \includegraphics[height=6.5cm,width=8.5cm]{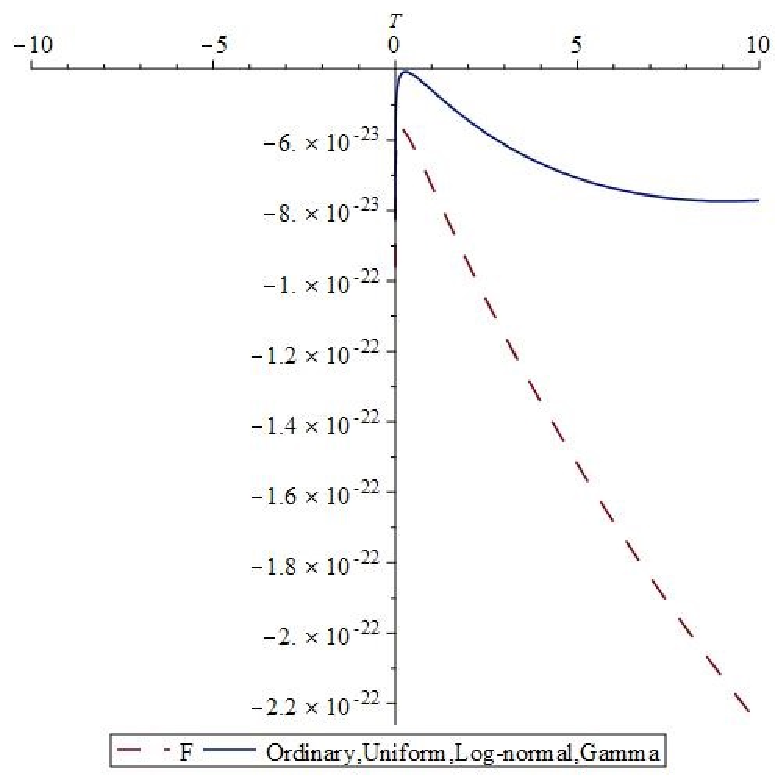}
 \label{6a}}
 \subfigure[]{
 \includegraphics[height=6.5cm,width=8.5cm]{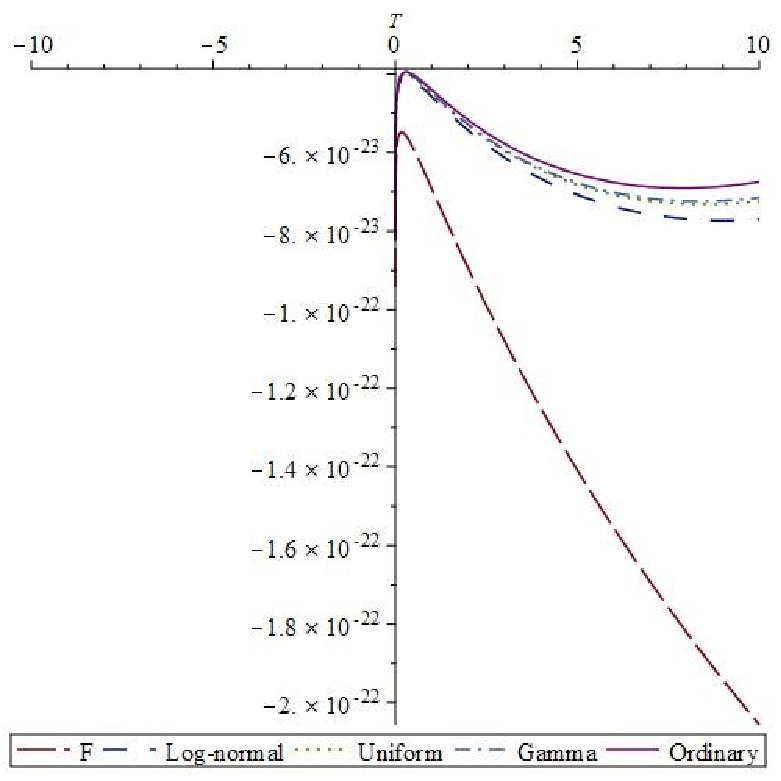}
 \label{6b}}
  \caption{\small{ Entropy in terms of temperature for different distributions with respect to $q=1, 1.1$ for figs 6a and 6b}}
 \label{6}
 \end{center}
 \end{figure}
\newpage

\begin{figure}[h!]
 \begin{center}
 \subfigure[]{
 \includegraphics[height=6.5cm,width=8.5cm]{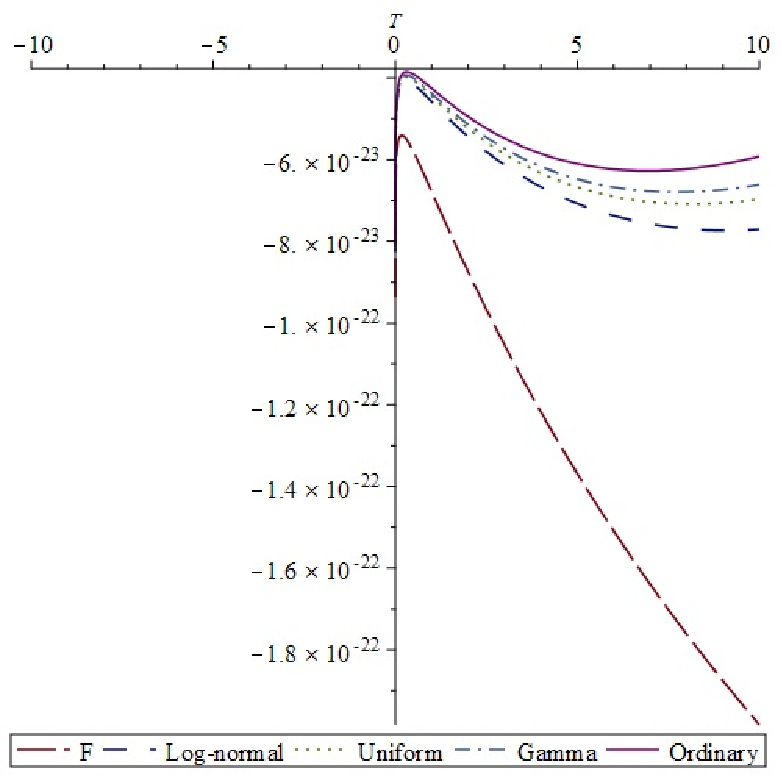}
 \label{7a}}
 \subfigure[]{
 \includegraphics[height=6.5cm,width=8.5cm]{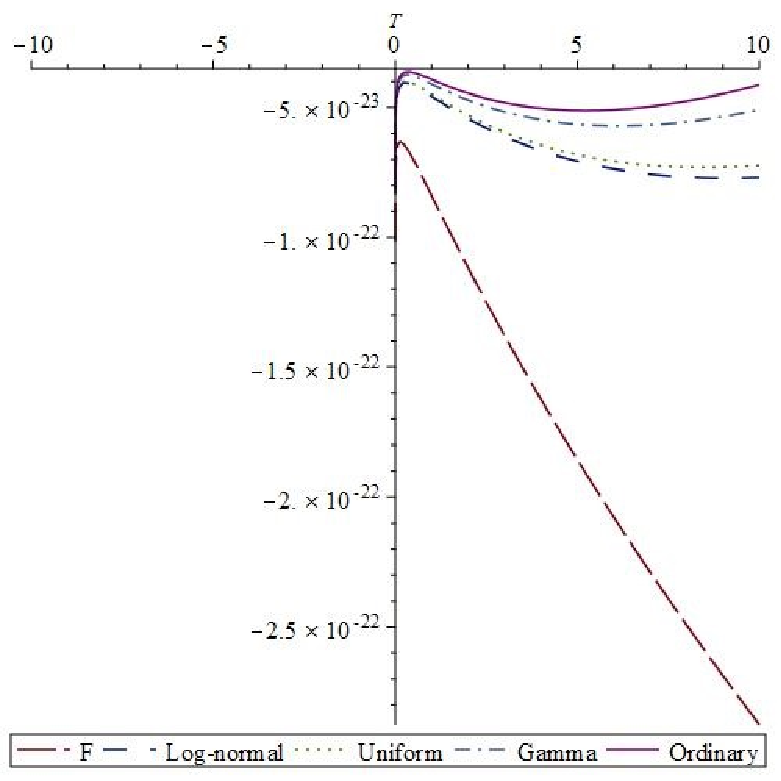}
 \label{7b}}
  \caption{\small{Entropy in terms of temperature for different distributions with respect to $q=1.2, 1.5$ for figs 7a and 7b}}
 \label{7}
 \end{center}
 \end{figure}
\newpage
  \begin{figure}[h!]
 \begin{center}
 \subfigure[]{
 \includegraphics[height=10.5cm,width=8.5cm]{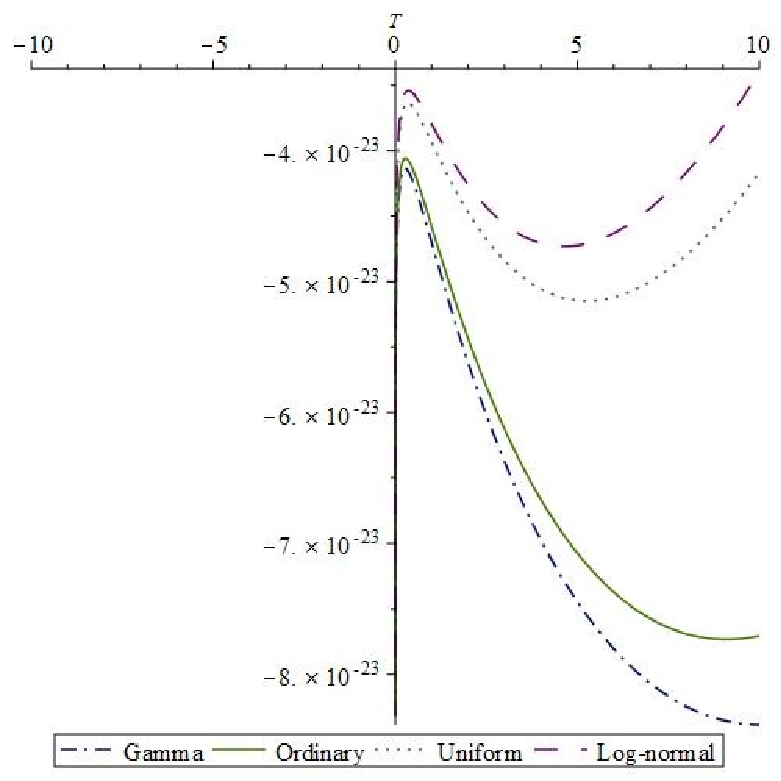}
 \label{8a}}
 \subfigure[]{
 \includegraphics[height=10.5cm,width=8.5cm]{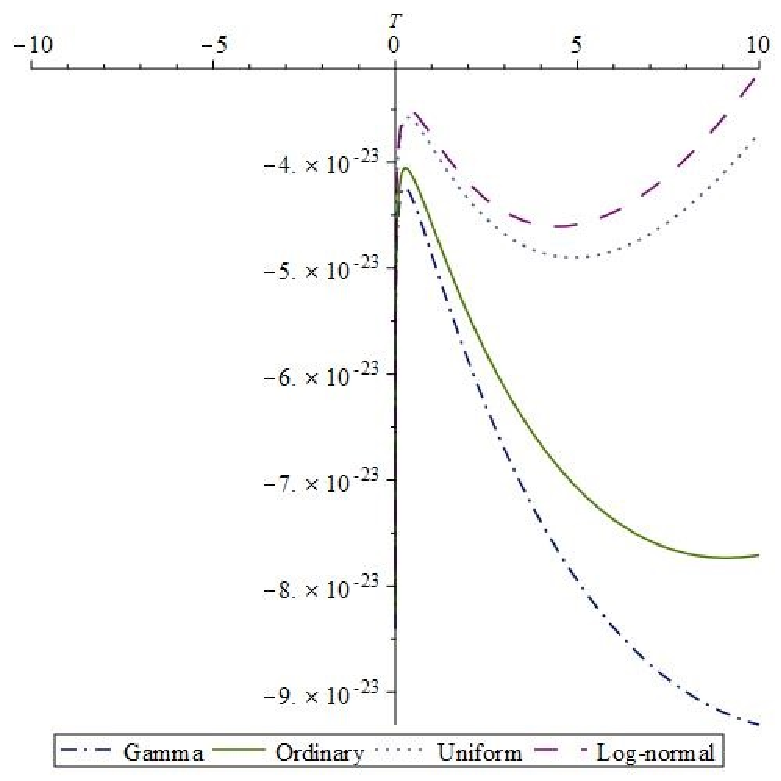}
 \label{8b}}
  \caption{\small{Entropy in terms of temperature for different distributions with respect to $q=1.7, 1.8$ for figs 8a and 8b}}
 \label{8}
 \end{center}
 \end{figure}
\newpage
   \begin{figure}[h!]
 \begin{center}
 \subfigure[]{
 \includegraphics[height=10.5cm,width=8.5cm]{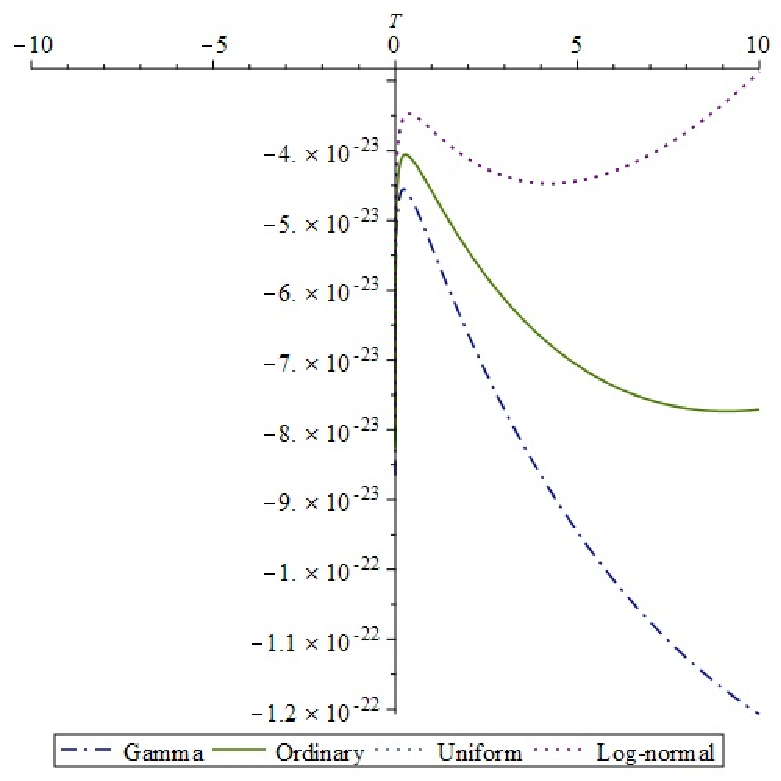}
 \label{9a}}
 \subfigure[]{
 \includegraphics[height=10.5cm,width=8.5cm]{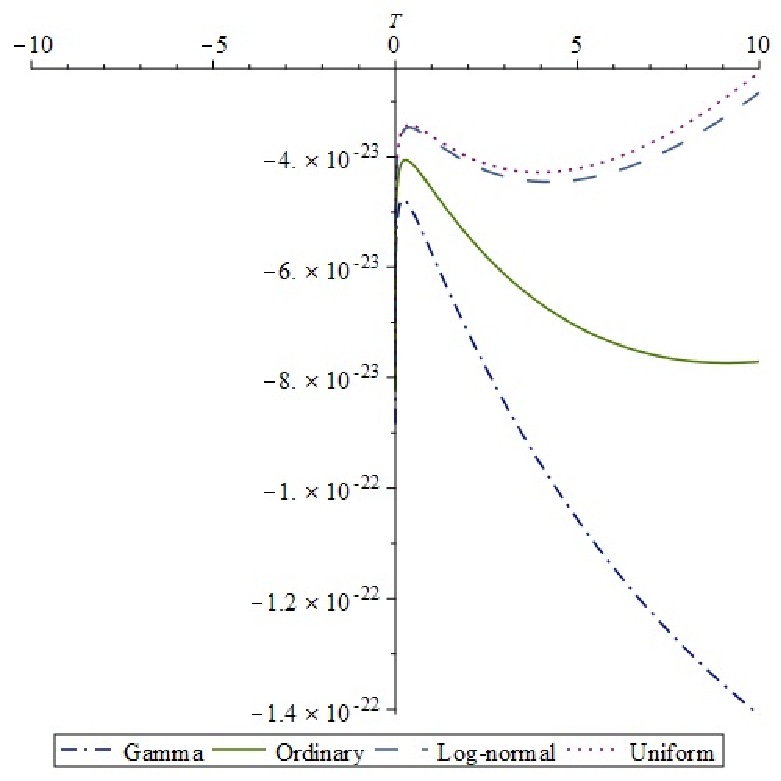}
 \label{9b}}
  \caption{\small{ Entropy in terms of temperature for different distributions with respect to $q=2, 2.1$ for figs 9a and 9b}}
 \label{9}
 \end{center}
 \end{figure}
\newpage

   \begin{figure}[h!]
 \begin{center}
 \subfigure[]{
 \includegraphics[height=10.5cm,width=8.5cm]{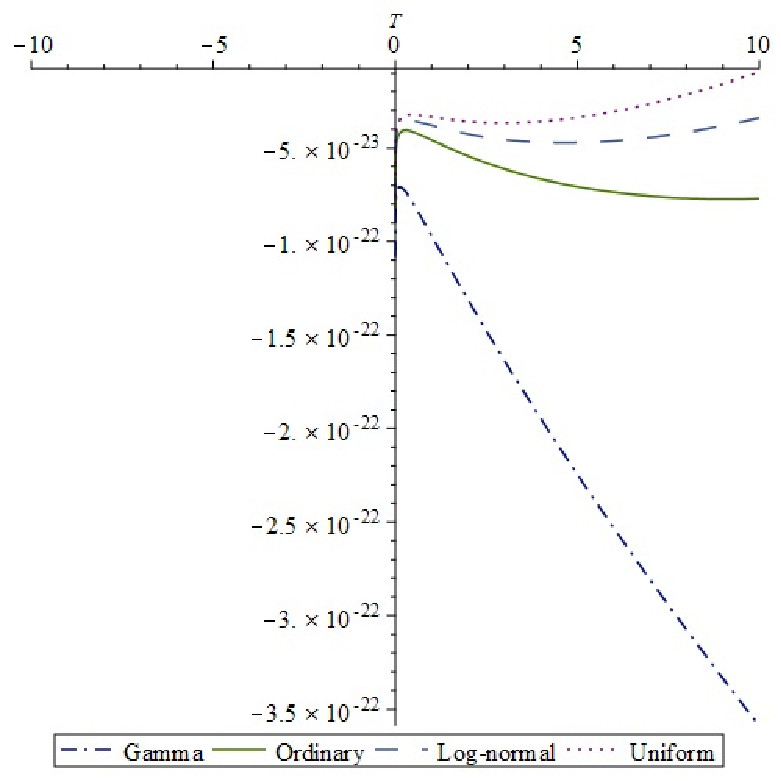}
 \label{10a}}
 \subfigure[]{
 \includegraphics[height=10.5cm,width=8.5cm]{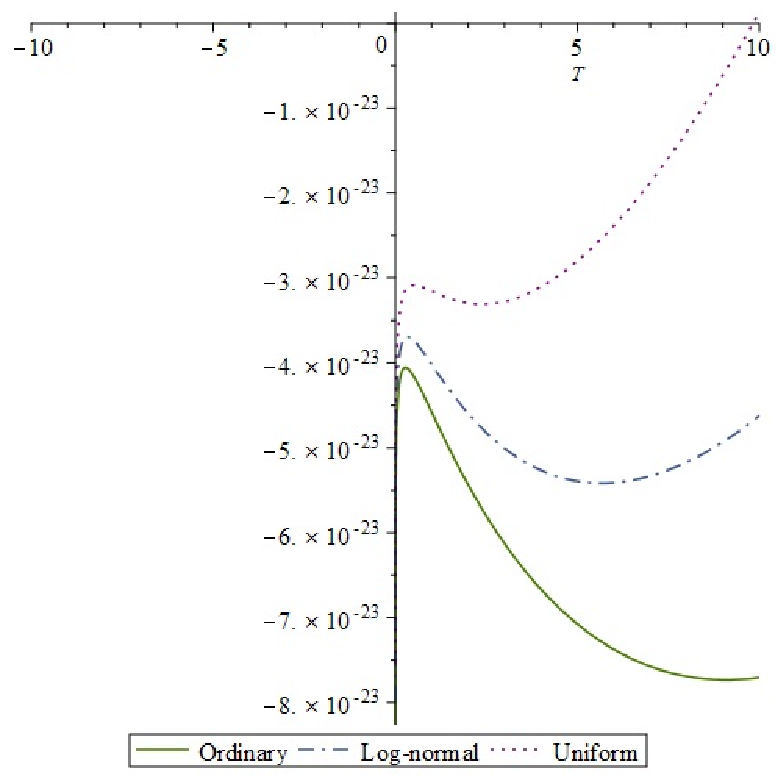}
 \label{10b}}
  \caption{\small{Entropy in terms of temperature for different distributions with respect to $q=2.5, 2.8$ for 10a and 10b}}
 \label{10}
 \end{center}
 \end{figure}

\end{document}